\documentclass[submission, Proceedings]{SciPost}

\hypersetup{
    colorlinks,
    linkcolor={red!50!black},
    citecolor={blue!50!black},
    urlcolor={blue!80!black}
}

\usepackage[bitstream-charter]{mathdesign}
\DeclareSymbolFont{usualmathcal}{OMS}{cmsy}{m}{n}
\DeclareSymbolFontAlphabet{\mathcal}{usualmathcal}
\newcommand{\sourav}{\textcolor{black}}
\newcommand{\caola}{\textcolor{black}}

\begin{document}

\begin{center}{\Large \textbf{
Eikonal Dressed Gluon Exponentiation to study power corrections to Thrust, $ C $-parameter, and Angularity
\\
}}\end{center}

\begin{center}
Neelima Agarwal\textsuperscript{1},
Ayan Mukhopadhyay\textsuperscript{2},
Sourav Pal\textsuperscript{2$\star$}, and
Anurag Tripathi\textsuperscript{2}
\end{center}

\begin{center}
{\bf 1} Department of Physics, Chaitanya Bharathi Institute of Technology,
Gandipet, Hyderabad, Telangana State 500075, India
\\
{\bf 2} Department of Physics, Indian Institute of Technology Hyderabad,
Kandi, Sangareddy, Telangana State 502285, India
\\
* Corresponding Author, spalexam@gmail.com
\end{center}

\begin{center}
\today
\end{center}


\definecolor{palegray}{gray}{0.95}
\begin{center}
\colorbox{palegray}{
  \begin{tabular}{rr}
  \begin{minipage}{0.1\textwidth}
    \includegraphics[width=35mm]{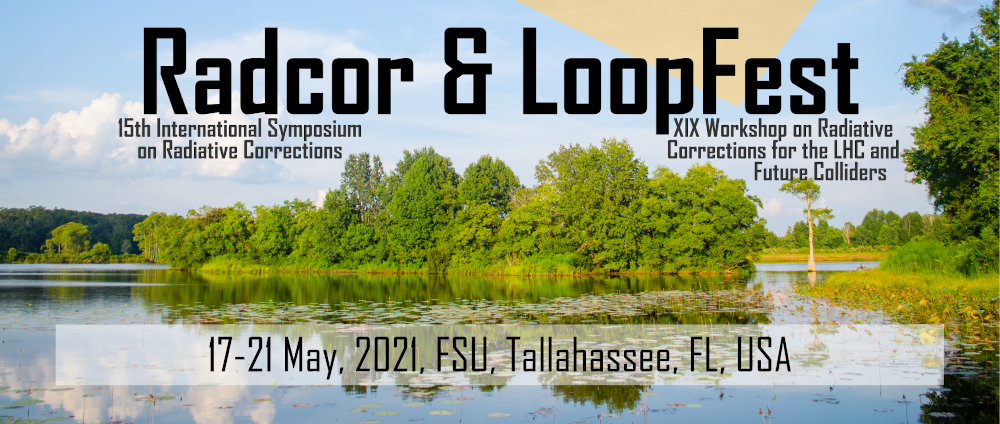}
  \end{minipage}
  &
  \begin{minipage}{0.85\textwidth}
    \begin{center}
    {\it 15th International Symposium on Radiative Corrections: \\Applications of Quantum Field Theory to Phenomenology,}\\
    {\it FSU, Tallahasse, FL, USA, 17-21 May 2021} \\
    \doi{10.21468/SciPostPhysProc.?}\\
    \end{center}
  \end{minipage}
\end{tabular}
}
\end{center}

\section*{Abstract}
{\bf
Dressed Gluon Exponentiation is a well known method to study power corrections. We present here its eikonalized version -- Eikonal Dressed Gluon Exponentiation (EDGE) \cite{Agarwal:2020uxi} to determine the dominant power corrections to shape variables such as Thrust, $ C $- parameter, and Angularity. Our method remarkably simplifies the calculations.   }

\vspace{10pt}
\noindent\rule{\textwidth}{1pt}
\tableofcontents\thispagestyle{fancy}
\noindent\rule{\textwidth}{1pt}
\vspace{10pt}

\section{Introduction}
\label{sec:intro}
Event shape variables are classical tools for precise determination of the strong coupling constant~\cite{Abbate:2010xh,Hoang:2015hka,Wang:2019isi,Marzani:2019evv,Gehrmann:2012sc} from collider data, and are useful while constructing the models for hadronization. Due to the infrared safety of the event shape variables, they can be calculated in the perturbation theory. The state of the art is NNLO calculations at fixed orders \cite{GehrmannDeRidder:2007hr,Weinzierl:2009ms,Gehrmann:2019hwf,Kardos:2020igb,GehrmannDeRidder:2009dp}, while NLL~\cite{Catani:1992ua,Banfi:2010xy,Banfi:2004yd,Banfi:2001bz}, and NNLL resummation \cite{Banfi:2016zlc,Becher:2012qc,Hoang:2014wka,Budhraja:2019mcz,Bell:2018gce,Hoang:2015gta,Kolodrubetz:2016reu,Lepenik:2019jjk,Lee:2009cw,Hornig:2009vb,Banfi:2014sua,Zhu:2021xjn} frameworks have been developed over several years. If one denotes the event shape variable by $ e $, then its corresponding distribution $ d\sigma/de $ peaks in the two jet limit ($ e\rightarrow0 $). In this particular limit the distributions are affected by the non-perturbative power corrections of order $ (\Lambda/eQ)^n $ and $ (\Lambda^2/eQ^2)^{n} $, arising from the soft gluon and the collinear gluon corrections respectively. \caola{Recently, linear power corrections for thrust, and $ C $-parameter were presented in \cite{Caola:2021kzt}}. Here, we discuss the analytic non-perturbative corrections, which are suppressed by powers of ($ \Lambda/Q $) as compared to the perturbative corrections.

A generic observable $ \sigma $ in perturbative QCD is a sum of perturbative corrections and the power corrections,
\begin{equation}
\sigma\left(\frac{Q}{\mu},\alpha_{s}\right)\,=\,\sigma_{{\rm pert}}\left(\frac{Q}{\mu_{f}},\frac{\mu_{f}}{\mu},\alpha_{s}\right)+\sum_{n}\sigma_{n}\left(\frac{\mu_{f}}{\mu},\alpha_{s}\right)\left(\frac{\mu_{f}}{Q}\right)^{n}\,,\label{gensig}
\end{equation}
where $ \mu_{f} $ is the factorization scale, and $ \mu $ is the renormalization scale. The perturbative part of the observable shows a factorial growth order by order in the strong coupling constant $ \alpha_{s} $, which implies that the perturbative series is divergent. One of the known techniques to sum a series with factorial growth is Borel summation, and upon performing Borel summation the factorial divergences appear as singularities in the Borel plane. We can avoid these singularities by contour deformation, however, the ambiguity in choosing the direction of the contour deformation results in ambiguous perturbative results. These ambiguities can be studied efficiently by studying the power corrections of these observables \cite{Beneke:1994sw}. For event shape distributions at LEP, the soft power corrections (corrections due to soft gluons) are dominant over the collinear power corrections (corrections due to collinear gluons). These soft power corrections are so dominant at the scale $ e\approx\Lambda/Q $ that we require resummation in order to make a stable prediction. However, the collinear corrections become relevant at smaller values of $ e\approx(\Lambda/Q)^2  $, which are generally out of experimental range. Thus, in this paper, we only discuss the soft power corrections.

An elegant and efficient way to deal with the perturbative logarithms 
and the non perturbative power corrections is Dressed gluon exponentiation (DGE)\cite{Gardi:2001di}. In this paper, we aim to calculate only the dominant soft power corrections using the Dressed gluon exponentiation. We merge the eikonal approximation for the emission of the soft gluons with the well known Dressed gluon exponentiation, and we call the resulting method eikonal Dressed gluon exponentiation (EDGE) \cite{Agarwal:2020uxi}. 

The paper is structured as follows: section \ref{sectionDGE} reviews the basic concepts about Dressed gluon exponentiation, section \ref{secEDGE} shows how EDGE can be used in terms of two different kinematic variables such as energy fractions and light cone co-ordinates to calculate the soft power corrections, and section \ref{expo} describes the calculation of Sudakov exponent by taking the Laplace transformation of the Borel function derived in section \ref{secEDGE}.   

\section{Dressed gluon exponentiation}  
\label{sectionDGE}  
The event shape distribution in the single dressed gluon approximation serves as the starting point for DGE, which is built from the one-loop real emission contribution to the event shape for a gluon with virtuality $ k^2\neq 0 $. The renormalon resummed event shape distribution, which dominates in the large $ N_f $ limit, is given by~\cite{Beneke:1994sw},
\begin{align}
\frac{1}{\sigma} \frac{d \sigma}{de} (e,Q^2) \, = \, - \, \frac{C_F}{2 \beta_0} 
\int_0^1  d \xi  \, \frac{d \mathcal{F} (e, \xi)}{d \xi} \, A(\xi Q^2) \, ,
\label{SDG}
\end{align}
where $\beta_0 = \frac{11}{12}C_A -\frac{1}{6} N_f$, $\xi = k^2/Q^2$, and $A(\xi Q^2)$ 
is the large-$\beta_0$ running coupling $(A = \beta_0 \alpha_s/\pi)$ on the time-like
axis. In the $\overline{\rm MS}$ scheme, the Borel representation of $ A(\xi Q^2) $ has the form,
\begin{align}
A(\xi Q^2) \, = \, \int_0^{\infty} du (Q^2/\Lambda^2)^{- u} 
\frac{\sin \pi u}{\pi u} {\rm e}^{\frac{5}{3}u} \xi^{- u} \, .
\label{largeb0coup}
\end{align}
\sourav{Now, if we denote the squared matrix element as $ \mathcal{M} $, then we can write down the characteristic function $ \mathcal{F}(e,\xi) $ as~\cite{Gardi:2000yh,Dokshitzer:1995qm}}, 
\begin{align}
\mathcal{F}(e, \xi) \, = \, \int dx_1 dx_2 \, {\cal M} (x_1, x_2, \xi) \, 
\delta \left( e -  \bar{e} (x_1, x_2, \xi) \right) \, ,
\label{charf}
\end{align}
\sourav{where $ x_i $'s are the energy fractions, and $\bar{e}$ is the expression of the event shape in terms of the kinematic variables. The Borel representation of cross-section distribution is obtained by interchanging the order of integration of eq. (\ref{SDG}), and it takes the form, }  
\begin{align}
\frac{1}{\sigma} \frac{d \sigma}{de} (e, Q^2) \, = \, \frac{C_F}{2 \beta_0} 
\int_0^{\infty} du (Q^2/\Lambda^2)^{- u} B(e,u) \, ,
\label{borrep}
\end{align}  
where the Borel function for single dressed gluon $B(e,u)$  is defined by, 
\begin{align}
B(e,u) \, = \, - \frac{\sin{\pi u}}{\pi u} e^{\frac{5}{3} u}
\int_{0}^{\infty} d \xi ~ \xi^{- u}  \frac{d \mathcal{F}(e,\xi)}{d \xi} \, .
\label{borfu}
\end{align}
This Borel function $ B(e,u) $ is free from any renormalon singularities in the $ u $ plane. However, when the Borel function for single dressed gluon is exponentiated via a Laplace transform, it generates the renormalon poles in the Borel plane, which we will discuss in section \ref{expo}.

\sourav{The exponentiation of the logarithmically enhanced terms in the Laplace space is guaranteed by the additive nature of the event shapes and the factorization of matrix elements for soft and collinear radiations. Then, the resummed cross-section in the Laplace space has the form, }
\begin{align}
\frac{1}{\sigma}\frac{d\sigma(e,Q^{2})}{de}=\int_{C-i\infty}^{C+i\infty}\frac{d\nu}{2\pi i}\,e^{\nu e}\,\text{exp}[S(\nu,Q^{2})] & ,\label{eq:resum}
\end{align}
where $C$ lies to the right of the singularities of the integrand. \sourav{This exponentiation in the Laplace space effectively resums the Sudakov logarithms and the power correction terms in the two-jet region}.
The Sudakov exponent has the form \cite{Gardi:2001ny}, 
\begin{align}
S(\nu,Q^{2}) & =\int_{0}^{1}de\:\frac{1}{\sigma}\frac{d\sigma(e,Q^{2})}{de}\:(e^{-\nu e}-1).\label{eq: sudakov-gen}
\end{align}
This integral is dominant in $\nu\rightarrow\infty$, which corresponds
to the Sudakov region $e\rightarrow0$. Using Eq. (\ref{borrep}), the
Sudakov exponent is reduced to,

\begin{align}
S(\nu,Q^{2})=\frac{C_{F}}{2\beta_{0}}\int_{0}^{\infty}du\,\left({Q^{2}}/{\Lambda^{2}}\right)^{-u}\,B_{\nu}^{e}(u),\label{eq:sudakov-gen-1}
\end{align}
where 
\begin{align}
B_{\nu}^{e}(u)=\int_{0}^{1}de\,B(e,u)\,(e^{-\nu e}-1).\label{eq: borel-int}
\end{align}
\section{Eikonal Dressed gluon exponentiation}  
\label{secEDGE}
In this section we describe Eikonal Dressed gluon exponentiation (EDGE) \cite{Agarwal:2020uxi} in two well known kinematic variables: energy fractions, and the light-cone variables. The basic idea of EDGE is to calculate the Borel function by considering the eikonal approximated matrix element and the eikonal approximated version of the relevant event shape variable. 
\subsection{EDGE in energy fractions}
\label{EDGEenergy}
In this subsection, we describe EDGE in the energy fraction variables $ x_i $.  \sourav{The colour and coupling
stripped off matrix element in the eikonal limit  for the process $\gamma^{*}\rightarrow q\bar{q}g$ with a massive gluon is given by}, 
\begin{align}
\mathcal{M}_{{\rm {soft}}}(x_{1},x_{2},\xi)=\frac{2}{(1-x_{1})(1-x_{2})}\,\label{eq:softmatr}
\end{align}
where,
\begin{align}
x_{1}= & \frac{2p_{1}\cdot Q}{Q^{2}},\quad x_{2}=\frac{2p_{2}\cdot Q}{Q^{2}},\quad x_{3}=\frac{2k\cdot Q}{Q^{2}},\qquad\xi=\frac{k^{2}}{Q^{2}}\,.\label{energyfrac}
\end{align}
Here $p_{1},p_{2}$, and $k$ denote the momenta of the quark, anti-quark,
and the gluon respectively, and $Q=p_{1}+p_{2}+k$, which fixes $ x_3=2-x_1-x_2 $. 
 
The eikonal approximated versions of thrust ($T$), $C$-parameter ($c$), and the angularities ($\tau_{a}$) are given by \cite{Agarwal:2020uxi}, 
\begin{align}
T\,=\,\text{Max}\left\lbrace x_{1},x_{2},\sqrt{x_{3}^{2}-4\xi}\right\rbrace, & \qquad 
c_{eik}(x_{1},x_{2})\,=\,\frac{(1-x_{1})(1-x_{2})}{(1-x_{1})+(1-x_{2})}, \nonumber \\ \nonumber\\
\tau_{a}^{eik}(x_{1},x_{2},\xi)\,=&\,(1-x_{1})^{1-a/2}(1-x_{2})^{a/2}.
\end{align}
Note that, the definition of thrust is so simple that it does not require any approximation in the eikonal limit, and we can use the full definition of thrust in EDGE. Also note that, we have used the modified definition \cite{Gardi:2003iv} of $C$-parameter, which is given by $ c=C/6 $.  Using the eikonal approximated event shapes and performing the integration over $ x_1,x_2 $  of eq. (\ref{charf}) by using the soft boundary of the phase space, we get the characteristic functions for thrust, $C$-parameter and angularities as, 
\begin{align}
\mathcal{F}\left(t,\xi\right)\,=&\,-\frac{4}{t}\text{log}\bigg(\frac{\xi}{t(q-t)}\bigg),\nonumber\\ \nonumber\\
\mathcal{F}(c,\xi)\,=&\,-\frac{4}{c} \text{log}\bigg(\frac{\xi}{c}\bigg) -8 \text{log}\bigg(1+\sqrt{1-\frac{4c^2}{\xi}}\bigg)-4\text{log}\bigg(\frac{1}{2}\bigg), \nonumber\\ \nonumber\\
\mathcal{F}(\tau_{a},\xi)\,=&\,-\frac{4}{\tau_{a}}\frac{1}{1-a}\log\xi.\label{Ftaupar3} \,
\end{align}
where $t=1-T$, and $q=\sqrt{T^{2}+4\xi}$. 
Now, using the results for characteristic functions and eq. (\ref{borfu}), we proceed for the calculation of the  Borel function for the emission of one gluon. The upper and the lower limit of the integration are determined from the collinear ($x_{1}=1-\xi,x_2=0$ ) and the soft ($x_{1}=x_{2}=1-\sqrt{\xi}$) gluon boundary conditions respectively. The Borel functions for thrust, $C$-parameter, and the angularities are given by \cite{Agarwal:2020uxi},
\begin{align}
B(t,u)\,= &\, \frac{\sin\pi u}{\pi u}e^{\frac{5u}{3}}\frac{4}{u}\frac{1}{t}\bigg(\frac{1}{t^{2u}}-\frac{1}{t^{u}}\bigg),\nonumber\\ \nonumber\\
B(c,u)\,=&\,4\frac{\text{sin}\pi u}{\pi u}e^{\frac{5u}{3}}\frac{1}{c}\bigg[\frac{1}{(2c)^{2u}}\frac{\sqrt{\pi}\Gamma(u)}{\Gamma(u+\frac{1}{2})}-\frac{1}{uc^{u}}\bigg],
\nonumber\\ \nonumber\\
B(\tau_{a},u)\,=&\,\frac{\sin{\pi u}}{\pi u}e^{\frac{5}{3}u}\frac{4}{1-a}\frac{1}{u}\frac{1}{\tau_{a}}\left[\frac{1}{\tau_{a}^{2u}}-\frac{1}{\tau_{a}^{\frac{2u}{2-a}}}\right]\,.\label{Bau}	
\end{align}
The results for the characteristics functions and the Borel functions for thrust, $C$-parameter and angularities match with the leading singular terms of the same function presented
in ~\cite{Gardi:2001ny} for thrust, ~\cite{Gardi:2003iv} for $C$-parameter, and ~\cite{Berger:2004xf} for angularities. 

Thus, we have obtained the correct leading singular terms using the eikonal definitions of the event shape variables and the soft approximated version of the squared matrix element, and as mentioned in ~\cite{Agarwal:2020uxi}, the computation of these functions is substantially less complex in EDGE as compared to the full calculation using the traditional Dressed gluon exponentiation. 
\subsection{EDGE in light-cone variables}
\label{sec:another}
In this subsection, we will calculate the Borel function for thrust, $C$-parameter,
and the angularities following the same steps of the previous subsection using the transverse momentum $k_{\perp}$ and rapidity $y$ of the massive gluon. In these variables, the eikonal approximated version of the three event shape variables considered in the previous section can be written as, 
\begin{align}
\bar{e}(k,Q)\,=\,\sqrt{\frac{k_{\perp}^{2}+k^{2}}{Q^{2}}}~h_{e}(y)\,,\label{class}
\end{align}
where $k_{\perp}$ and $y$ denote transverse momentum of the gluon
and pseudo-rapidity measured with respect to the thrust axis respectively. For thrust ($t$), $C$-parameter ($c$), and angularities($\tau_a$), $ h_e(y) $ takes the form,
\begin{align}
h_{t}(y)\,=&\,e^{-|y|} \, , \qquad 
h_{c}(y)\,=\,\frac{1}{2\cosh y} \, , \qquad 
h_{\tau_{a}}(y)\,=\, e^{-|y|(1-a)}\, .
\label{def-in-rapidity}
\end{align}
The matrix element for the emission of soft gluons can be computed easily by applying the eikonal Feynman rules. The fact that the soft phase space factorizes from the hard part reduces the soft cross-section to a very simple universal form, 
\begin{equation}
\frac{d\sigma}{\sigma}\,=\,\frac{1}{3}\,\frac{4}{k^{2}+k_{\perp}^{2}}\,dk_{\perp}^{2}dy\,.
\label{eq:cross-section--lighcone}
\end{equation}
Now, using eq. (\ref{charf}) and (\ref{eq:cross-section--lighcone}) and integrating over the transverse momentum $ k_{\perp} $, the characteristics function takes a very simple form \cite{Agarwal:2020uxi},
\begin{align}
\ensuremath{\mathcal{F}}(e,\xi)\,=\,\frac{8}{e}\int_{y_{{\rm min}}}dy\,.\label{uniF2}
\end{align}  
Here, only the lower limit is relevant for soft power corrections. The lower limit for this integration is  calculated  using eq. (\ref{class}) and (\ref{def-in-rapidity}), and for three event shape variables they are given by, 
\begin{align}
y_{{\rm min}}(t)\,=\,\log\,\Bigg(\frac{1}{t}\sqrt{\xi}\Bigg)\,,&\quad
y_{{\rm min}}(c)\,=\,\cosh^{-1}\Bigg({\sqrt{\xi}}/2c\Bigg),\nonumber \\ \nonumber \\
y_{{\rm min}}(\tau_{a})\,=&\,\frac{1}{1-a}\,\log\,\Bigg(\frac{1}{\tau_{a}}\sqrt{\xi}\Bigg)\,.
\label{ymin}
\end{align}
Using eq.  (\ref{uniF2}) and (\ref{ymin}), we get the characteristics functions for the thrust, $C$-parameter, and Angularities as, 
\begin{align}
\mathcal{F}(t,\xi)\,=\,-\frac{8}{t}\,\text{log}\,\Bigg(\frac{\sqrt{\xi}}{t}\Bigg)\,, & \quad
\mathcal{F}(c,\xi)\,=\,-\frac{8}{c}\,\text{cosh}^{-1}\,\Bigg(\frac{\sqrt{\xi}}{2c}\Bigg)\, ,\nonumber \\ \nonumber \\
\mathcal{F}(\tau_a,\xi)\,=&\,-\frac{8}{\tau_{a}(1-a)}\log\Bigg(\frac{\sqrt{\xi}}{\tau_{a}}\Bigg)\,,
\end{align} 
which again agrees with the known results presented in \cite{Gardi:2001ny,Gardi:2003iv,Berger:2004xf} for the three event shape variables.
Now, using eq. (\ref{borfu}) one can calculate the Borel functions, which will be same as those presented in  eq. (\ref{Bau}). 
\section{Exponentiation}
\label{expo}
In this section, we will compute the Sudakov exponent for thrust. Similar conclusions can be made for the other two shape variables mentioned in the article.  
\sourav{In the Sudakov region ($ \nu \rightarrow \infty $), using $ B(t,u) $ of eq. (\ref{Bau}), and keeping only the logarithmic enhanced terms, the Borel function for thrust in Laplace space can be written as, }
\begin{align}
B_{\nu}^{t,\text{{eik}}}(u)=2\,e^{\frac{5}{3}u}\frac{\,\text{sin\,}\pi u}{\pi u}\bigg[\Gamma(-2u)\:\bigg(\nu^{2u}-1\bigg)\frac{2}{u}-\Gamma(-u)\:\bigg(\nu^{u}-1\bigg)\frac{2}{u}\bigg] & .\label{eq: borel-thrust-fin}
\end{align}
\sourav{Note that, this is free from any $ u=0 $ singularities. The coefficient of $ \Gamma(-2u) $ corresponds to large-angle soft gluon emissions, while the coefficient of $ \Gamma(-u) $ corresponds to the collinear gluon emissions. However, this expression contains two types of poles: $\Gamma(-2u) $ has poles for all positive integers and half-integers, and $ \Gamma(-u) $ has poles for all positive integers. The prefactor $\sin \pi u$ regulates the poles for the integer values of $u$.} 

We will now compare our result with the full result for \textbf{$B_{\nu}^{t}$}
presented in \cite{Berger:2004xf,Gardi:2003iv} which\textbf{ }is
given by, 
\begin{align}
B_{\nu}^{t}(u)=\,2{\rm e}^{\frac{5}{3}u}\,\frac{\sin\pi u}{\pi u}\left[\Gamma(-2u)\left(\nu^{2u}-1\right)\frac{2}{u}-\Gamma(-u)\left(\nu^{u}-1\right)\left(\frac{2}{u}+\frac{1}{1-u}+\frac{1}{2-u}\right)\right]\,.\label{eq: Btfull}
\end{align}
We note that $B_{\nu}^{t,\text{eik}}(u)$ does not contain $ u=1 $ and $ u=2 $ poles as compared to the full $ B_{\nu}^{t}(u) $.
\sourav{We note that the poles related to the large angle soft gluon emissions are same in the approximated and the full result, thus leading logarithm terms are same for the full $ B_{\nu}^{t}(u) $ and $ B_{\nu}^{t,\text{eik}}(u) $. A detailed comparison of the leading and the sub-leading logarithms are discussed in \cite{Agarwal:2020uxi}. It is important to note that EDGE does not produce any new spurious renormalon poles.} 

\sourav{The poles in $B_{\nu}^{t}(u) $ in the real $ u $-axis implies that the integral over $ u $ to determine the Sudakov exponent of eq. (\ref{eq:sudakov-gen-1}) is ill-defined and one needs to deform the contour to evaluate this integral. The deformation of the contour around the real $ u $-axis introduces an ambiguity in the result, which can be determined by calculating the residue of $B_{\nu}^{t}(u) $ at the poles. The sizes of these residues determine the amount of the power corrections. It is evident from the expression of $B_{\nu}^{t}(u) $, that the power corrections due to soft gluons will arise from the poles at $ u=m/2 $, whereas the collinear power corrections will arise from the poles at $ u=1 $, and $ u=2 $. It was discussed in \cite{Agarwal:2020uxi,Berger:2004xf,Gardi:2001ny,Gardi:2003iv} that the soft power corrections are dominant over the collinear corrections. For example, at LEP, the collinear power corrections are 1000 times smaller than soft corrections \cite{Agarwal:2020uxi}. As EDGE correctly reproduces the poles related to the emission of large angle soft gluons, thus we can determine the soft power corrections correctly using EDGE.} 
\sourav{The dominance of the soft power corrections is true for all the event shapes considered in this paper. Thus the dominant soft power correction for other variables considered in this paper can be calculated using the same steps.} 

\section{Conclusion}
\sourav{In this paper, we have discussed EDGE, developed in \cite{Agarwal:2020uxi}, based on the well known Dressed gluon exponentiation. We have demonstrated that the soft power corrections can be calculated using the eikonal squared matrix element, together with the eikonal version of the relevant event shape variable. We have also observed that EDGE does not generate any new spurious renormalon poles in the Sudakov exponent. We believe that, as compared to traditional Dressed Gluon Exponentiation, EDGE  is simple enough and can be extended to the hadronic event shape variables, which are relevant at the LHC.} 

\section*{Acknowledgements}
SP would like to thank the organisers of RADCOR-LoopFest 2021. SP and AM would like to thank MHRD Govt. of India for an SRF fellowship.
AT would like to thank Lorenzo Magnea for suggesting this project
and for very fruitful discussions, Einan Gardi for very useful discussions,
and the University of Turin and INFN Turin for warm hospitality during
the course of this work. 
\bibliography{fullref}
\nolinenumbers
\end{document}